\documentclass{aa}  
\usepackage{graphicx}
\usepackage{txfonts}

\usepackage{natbib}
\usepackage{amsmath} 
\usepackage{amssymb}
\usepackage{mathtools}
\usepackage{microtype}
\usepackage[dvipsnames]{xcolor}
\definecolor{CiteRed}{RGB}{110, 0, 0}
\usepackage{multirow}
\usepackage{array}
\usepackage{soul} 
\usepackage{makecell}
\usepackage{siunitx}
\usepackage{xcolor}     
\usepackage{orcidlink}  
\usepackage{graphicx}
\usepackage{subcaption}
\usepackage{verbatim}
\graphicspath{ {img/} {plot/} {fig/} }
\defcitealias{SternSvensson96}{SS96}

\usepackage{hyperref}
\hypersetup{
    breaklinks,
    colorlinks,
    citecolor=CiteRed,  
    linkcolor=NavyBlue, 
    urlcolor=NavyBlue,  
    linktoc=page,
    pdftitle={Long gamma--ray burst light curves as the result of a common stochastic pulse--avalanche process},
    pdfkeywords={GRBs, Machine Learning, Genetic Algorithms},
    pdfauthor={Lorenzo Bazzanini},
    pdfcreator={\LaTeX}
}

\begin{document} 

   \title{Long gamma--ray burst light curves as the result of a common stochastic pulse--avalanche process}

   \author{L.~Bazzanini~\inst{1,2}\fnmsep\thanks{\texttt{bzzlnz[at]unife[dot]it}}\orcidlink{0000-0003-0727-0137}\and
           L.~Ferro~\inst{1,2}\orcidlink{0009-0006-1140-6913}\and
           C.~Guidorzi~\inst{1,2,3}\orcidlink{0000-0001-6869-0835}\and
           G.~Angora~\inst{1,4}\orcidlink{0000-0002-0316-6562}\and
           L.~Amati~\inst{2}\orcidlink{0000-0001-5355-7388}\and
           M.~Brescia~\inst{4,5}\orcidlink{0000-0001-9506-5680}\and 
           M.~Bulla~\inst{1,3,6}\orcidlink{0000-0002-8255-5127}\and
           F.~Frontera~\inst{1,2}\orcidlink{0000-0003-2284-571X}\and
           R.~Maccary~\inst{1}\orcidlink{0000-0002-8799-2510}\and
           M.~Maistrello~\inst{1}\orcidlink{0009-0000-4422-4151}\and
           P.~Rosati~\inst{1,2,3}\orcidlink{0000-0002-6813-0632}\and
           A.~Tsvetkova~\inst{7,2,8}\orcidlink{0000-0003-0292-6221}
          }
          
   \institute{
   Department of Physics and Earth Science, University of Ferrara, via Saragat 1, I--44122, Ferrara, Italy\label{unife}\and 
   INAF -- Osservatorio di Astrofisica e Scienza dello Spazio di Bologna, Via Piero Gobetti 101, I-40129 Bologna, Italy\label{oabo}\and 
   INFN -- Sezione di Ferrara, via Saragat 1, I--44122, Ferrara, Italy\label{infnfe}\and
   INAF -- Osservatorio Astronomico di Capodimonte, Salita Moiariello 16, I-80131 Napoli, Italy\label{oacn}\and
   Dipartimento di Fisica ``E. Pancini'', Universit\`a di Napoli ``Federico II'', Via Cinthia 21, I-80126 Napoli, Italy\and
   INAF, Osservatorio Astronomico d’Abruzzo, via Mentore Maggini snc, 64100 Teramo, Italy\and
   Department of Physics, University of Cagliari, SP Monserrato-Sestu, km 0.7, 09042 Monserrato, Italy\and
   Ioffe Institute, Politekhnicheskaya 26, 194021 St. Petersburg, Russia\\
   }

  \date{Received xxx; accepted xxx}

  \abstract
   {The complexity and variety exhibited by the light curves of long gamma--ray bursts (GRBs) enclose a wealth of information that still awaits being fully deciphered. Despite the tremendous advance in the knowledge of the energetics, structure, and composition of the relativistic jet that results from the core collapse of the progenitor star, the nature of the inner engine, how it powers the relativistic outflow, and the dissipation mechanisms remain open issues.}
   {A promising way to gain insights is describing GRB light curves as the result of a common stochastic process. In the Burst And Transient Source Experiment (BATSE) era, a stochastic pulse avalanche model was proposed and tested through the comparison of ensemble-average properties of simulated and real light curves. Here we aim to revive and further test this model.}
   {We apply it to two independent data sets, BATSE and {\it Swift}/BAT, through a machine learning approach: the model parameters are optimised using a genetic algorithm.}
   {The average properties are successfully reproduced. Notwithstanding the different populations and passbands of both data sets, the corresponding optimal parameters are interestingly similar. In particular, for both sets the dynamics appears to be close to a critical state, which is key to reproduce the observed variety of time profiles.}
   {Our results propel the avalanche character in a critical regime as a key trait of the energy release in GRB engines, which underpins some kind of instability.}

   \maketitle
    

\section{Introduction}
\label{sec:intro}

Gamma--ray bursts (GRBs) are the most powerful explosions on stellar scale in the Universe. At least two kinds of progenitors are known: (a) so-called `collapsar'~\citep{Woosley93, Paczynski98, MacFadyen99}, that is a hydrogen-stripped massive star, whose core collapses to a compact object, which launches a relativistic ($\Gamma\sim 10^2$--$10^3$) jet; (b) merger of a compact binary~\citep{Eichler89, Paczynski91, Narayan92}, where at least one of the two objects is supposed to be a neutron star (NS) and which also results in a short-lived relativistic jet (see~\citealt{KumarZhang15rev, Zhang18_book} for recent reviews). There are alternative models to the collapsar, such as the binary-driven hypernova model (BdHN; \citealt{RuedaRuffini12, Becerra19}), in which the final collapse of a CO core of a massive star can trigger the collapse of a companion neutron star. Most GRBs with progenitor of type (a) manifest themselves as long GRBs (LGRBs), lasting longer than $\sim \SI{2}{s}$\footnote{This boundary value is from CGRO/BATSE GRB catalogue and slightly depends on the detector's passband.}, while (b) usually exhibit a subsecond spike occasionally followed by weak, long-lasting emission~\citep{Norris06}, and are commonly referred to as short GRBs (SGRBs). Actually, the emerging picture is more complicated, as shown by the increasing number of cases with deceptive time profiles found in both classes~\citep{Gehrels06, Rastinejad22, Gompertz23, Yang22, Troja22, Ahumada21, Zhang21a, Rossi22a, Levan23, Levan24}.

The nature of the dissipation mechanism that is responsible for the GRB prompt emission is still an open issue. The great variety observed in the light curves (LCs) of LGRBs is thought to be the result of the variability imprinted to the relativistic outflow by the inner engine left over by the collapsar, either a millisecond magnetised NS or a black hole (BH), along with the effects of the propagation of the jet within the stellar envelope (e.g., \citealt{Morsony10, Geng16, Gottlieb20a, Gottlieb20b, Gottlieb21a, Gottlieb21b}), although some models ascribe the possible presence of subsecond variability to magnetic reconnection events taking place at larger radii~\citep{ICMART}. Some correlations were found between variability and minimum variability timescale on one side, as defined in a number of ways, and luminosity and initial Lorentz factor of the outflow on the other side (e.g., see~\citealt{Camisasca23} and references therein). However, apart from the study of average and of individual Fourier power density spectra of GRBs~\citep{Beloborodov98, Guidorzi12, Guidorzi16, Dichiara13a}, the study of the waiting time distribution between pulses~\citep{RamirezRuiz01b, NakarPiran02a, Quilligan02, Guidorzi15b}, and the distribution of the number of peaks per GRB~\citep{Guidorzi24}, little progress has been made in deciphering and characterising the variety of LGRB LCs within a unifying scheme that could explain the large diversity (in terms of duration, number of pulses, distribution of energy and waiting times between pulses) and relate it to other key properties. 

Given the observed erratic behaviour of GRB LCs, two primary hypotheses can be identified: (i) the apparent lack of periodicity or distinct patterns, which precludes an easy classification and the prediction of GRB LCs, may originate from deterministic, non-linear, and possibly chaotic processes; (ii) the energy release mechanism might be inherently stochastic. While hypothesis (i) has been explored in previous studies (e.g., \citealt{Greco11}) and remains plausible, our work focuses instead on scenario (ii). A stochastic character is also naturally invoked in the presence of turbulence. For instance, magnetised GRB outflows can experience turbulent magnetic reconnection episodes~\citep{lazarian1999}.

Recent investigations found possible evidence that GRB engines emit as self-organised critical (SOC) systems~\citep{WangDai13, Yi17, Lyu20, Wei23, Li23b, Maccary24}, in which energy is released through avalanches whenever the system naturally reaches a critical point. Yet, the interpretation is not straightforward, since SOC dynamics is usually invoked for systems that are continuously fed by some energy input and are not characterised by the kind of irreversible evolution expected for a GRB inner engine. A successful description of the inner engine variability would help to constrain the mechanism that powers the jetted outflow in GRBs and, ultimately, the nature of the compact object. Furthermore, it would provide the community with a reliable tool to simulate credible GRB LCs as they would be measured by future experiments, avoiding the pitfalls of using real noisy LCs (e.g., \citealt{Sanna20}).

In this respect, an interesting attempt was laid out by~\citet[hereafter SS96]{SternSvensson96} in the Compton Gamma--Ray Observatory era (CGRO; 1991--2000) on the GRB catalogue of one of its experiments, the Burst And Transient Source Experiment (BATSE).
These authors proposed a common stochastic process built on a pulse avalanche mechanism and tried to reproduce some of the observed distributions of BATSE GRB LCs. At that time, the cosmological distances of GRBs and the progenitors' nature of the two classes were yet to be firmly established, with the first afterglow discoveries starting from 1997~\citep{Costa97}.
By manually guessing the values of the seven model parameters, \citetalias{SternSvensson96} came up with a process operating in a nearly critical regime and capable of reproducing the variety of observed GRB LCs, as long as the chosen metrics are concerned. This approach of simulating GRB LCs was adopted in~\citet{Greiner2022} to assess localisation capabilities of a proposed network of GRB detectors on the global navigation satellite system Galileo G2.

The~\citetalias{SternSvensson96} model represents an interesting attempt to capture essential aspects of the way energy is dissipated over time beyond a generic and rather uninformative attribute such as ``random'' used to describe the sequence of peaks and pulses.
The core mechanism of this model is the avalanche through which energy is released as a runaway process triggered by some kind of instability: some initial episodes of energy release, which turn up as major pulses in the LC (so-called ``parent pulses''), can further give rise to ``child pulses'' and so on, until the process becomes subcritical.

A possibility that could give rise to an avalanche process like the one underpinning the~\citetalias{SternSvensson96} model is given by magneto-rotational and/or gravitational instabilities in the hyperaccreting disc of the left-over compact object, which then cause fragmentation (also proposed by~\citealt{Perna06} to explain GRB X-ray flares). Child pulses could result from further fragmentation of a parent episode, or viscous instabilities that form clumps, which can imprint a branching character (e.g., \citealt{Kawanata13, Shahamat21}), or current-driven instabilities as in the Poynting-flux dominated outflow by~\citet{LyutikovBlandford03}. Another possibility is related to the runaway character of the magnetic reconnection events that are envisaged in the Internal-Collision-induced MAgnetic Reconnection and Turbulence model (ICMART; \citealt{ICMART}).

In the big data era, advanced statistical and machine learning (ML) techniques applied to astrophysics have become routine (e.g., see~\citealt{Feigelson21} for a review). In this paper we aim to verify and improve the results obtained by~\citetalias{SternSvensson96} on the BATSE data and, for the first time, apply their model to a sample from another detector operating in a softer energy band, such as the Burst Alert Telescope (BAT;~\citealt{Barthelmy05}) aboard
the {\it Neil Gehrels Swift} Observatory~\citep{Gehrels04}. Specifically, we aim at optimising the model parameters using a genetic algorithm (GA; \citealt{rojas1996}). 

A similar technique, in which the parameters of a physical model were optimised through the application of a GA, was recently applied  by~\citet{vargas2022survival} to model the shock propagation in the supernova SN2014C progenitor star and ejecta, in which the GA was used to optimise a hydrodynamic and radiation transfer model.

Unlike~\citetalias{SternSvensson96}, we restrict our analysis to LGRBs, whose progenitor is thought to be a collapsar, to preserve as much as possible the homogeneity of the putative GRB inner engines. For $\sim30$\% of the {\it Swift} sample with measured redshift, in principle it is possible to carry out the same analysis in the GRB rest frame. However, we did not consider this option, since the cosmological dilation correction by $(1+z)$ is partly counteracted by other energy-dependent effects, which make the final correction milder and less obvious (see~\citealt{Camisasca23} and references therein for a detailed explanation).

In this paper, we report the main results and implications. A companion and more ML-oriented paper will report all the technical details. The present work is organised as follows: in Section~\ref{sec:data} we describe the data analysis and sample selection, while in Section~\ref{sec:met} we illustrate the methods underpinning the avalanche model and the implementation of the genetic algorithm. Section~\ref{sec:res} reports the results, whose discussion and conclusions are laid out in Section~\ref{sec:disc_conc}.


\section{Data Analysis}
\label{sec:data}

\subsection{Sample selection}\label{ss:lc_selection}

From the BATSE 4B catalogue~\citep{Paciesas99} we took the 64-ms time profiles that were made available by the BATSE team\footnote{\url{https://heasarc.gsfc.nasa.gov/FTP/compton/data/batse/ascii_data/64ms/}}. Observed with the BATSE eight Large Area Detectors (LADs), these data are the result of a concatenation of three standard BATSE types, DISCLA, PREB, and DISCSC, available in four energy channels: 25--55, 55--110, 110--320, and $>\SI{320}{keV}$. We used the total passband LCs.
For each GRB the background was interpolated with polynomials of up to fourth degree as prescribed by the BATSE team. In our analysis, we used the background-subtracted LCs.

From an initial sample of 2024 GRBs we selected only those that satisfy the following requirements:
\begin{itemize}
    \item $T_{90} > \SI{2}{s}$, that is, only \textit{long} GRBs;
    \item data available for at least $\SI{150}{s}$ after the brightest peak; 
    \item signal-to-noise ratio (S/N) of the total net counts within the duration of the event greater than $70$.
\end{itemize}
In order to estimate the S/N, following~\citetalias{SternSvensson96}, rather than the commonly used $T_{90}$, we used as a proxy of the GRB duration the time interval from the first to the last time bin whose counts exceed the threshold of $20\%$ of the peak counts, henceforth called $T_{20\%}$. Before evaluating the $T_{20\%}$, the LCs were first convolved with a Savitzky-Golay smoothing filter~\citep{1964AnaCh..36.1627S}, using a second order interpolating polynomial, and a moving window of size $T_{90}/15$. Accordingly, we defined the S/N of a GRB as the sum of the net counts in the whole $T_{20\%}$ interval, divided by the corresponding error. The value of the S/N threshold was the result of a trade-off between the number of GRBs and the statistical quality of the LCs in the sample. Furthermore, the $T_{20\%}$ is also used to compute the duration distribution of the LCs (Section~\ref{sec:metrics}).

We ended up with 585 long GRBs satisfying the aforementioned properties. Hereafter, this will be referred to as the BATSE sample.

As a second dataset, we considered the GRBs detected by {\it Swift}/BAT from January 2005 to November 2023 and covered in burst mode. We used the total 15--150~keV passband LCs, with 64-ms bin time; these were extracted as mask-weighted background-subtracted LCs, following the standard procedure recommended by the BAT team.\footnote{\url{https://swift.gsfc.nasa.gov/analysis/threads/bat\_threads.html}.}
From an initial sample of 1389 GRBs observed in burst mode, 531 passed the selection based on the same criteria adopted for BATSE, except for the value of the S/N threshold, which was lowered to 15 to obtain a sample of comparable size to the BATSE one, but still ensuring the required statistical quality. Hereafter, this will be referred to as the \textit{Swift} sample.

\subsection{Statistical metrics}
\label{sec:metrics}
We considered the following four metrics, which were also used by~\citetalias{SternSvensson96}: 
\begin{enumerate}
    \item the average peak-aligned post-peak time profile~\citep{1996MmSAI..67..417M}, in the time range $\numrange{0}{150}\,\si{s}$ \textit{after} the brightest peak. It is evaluated by averaging the normalised count rate of all the LCs in the sample, i.e.~$\langle F / F_p\rangle$, $F_p$ being the peak count rate. Further details are given in~\citet{stern1996b}; 
    \item the average peak-aligned third moment of post-peak time profiles $\langle(F / F_p)^3\rangle$, evaluated analogously to the first moment; 
    \item the average auto-correlation function (ACF). For both data samples the ACF is corrected for the counting statistics noise as in~\citet{link93acf} and is computed in the $0$--$150$~s interval; 
    \item the $T_{20\%}$ distribution, with $T_{20\%}$ used as a proxy of the duration.
\end{enumerate}
As in~\citetalias{SternSvensson96}, (1)--(4) are used as metrics to evaluate the degree of similarity between the real and the simulated LCs.


\section{Methods}
\label{sec:met}

\subsection{Light curve simulations}
\label{sec:ss96model}
 The stochastic process conceived by~\citetalias{SternSvensson96} belongs to the class of so-called `branching' processes, which describe the development of a population whose members reproduce according to some random process~\citep{Harris_branching_processes}. The process by~\citetalias{SternSvensson96} aims to mimic the temporal properties of GRBs and works as a \textit{pulse avalanche}, which is a linear Markov process. We outline its key features below, and refer the reader to~\citetalias{SternSvensson96} for more details.

The~\citetalias{SternSvensson96} model is based on the assumptions that:
\begin{enumerate}
    \item GRB LCs can be viewed as distinct random realisations of a common stochastic process, within narrow parameter ranges;
    \item the stochastic process should be scale invariant in time;
    \item it operates close to a critical state.
\end{enumerate}
With this model, each LC consists of a series of spontaneous primary (or {\it parent}) pulses, each of which can give rise to secondary (or {\it child}) pulses, which can then recursively spawn other generations of child pulses until the process reaches subcritical conditions and stops. Each pulse, which acts as a building block, is described by a Gaussian rise followed by a simple exponential decay:
\begin{equation}
    f(t) =
    \begin{cases}
        A \exp \left\{-(t-t_p)^2/\tau_r^2\right\}, & \text{for $t < t_p$}\\
        A \exp \left\{-(t-t_p)/\tau\right\},       & \text{for $t > t_p$}
    \end{cases}\quad,
    \label{eq:pulse}
\end{equation}
where $\tau$ is roughly the pulse width, $t_p$ is the peak time, $A$ is the amplitude, and we assume $\tau_r=\tau/2$~\citep{Norris96}. 
Differently from~\citetalias{SternSvensson96}, we do not sample $A$ from a uniform distribution $\mathcal{U}[0,1]$, rather, for each GRB we sample the value $A_{\max}$ from the distribution of the peak count rates of the real observed LCs, and then the amplitude of each pulse composing that LC is sampled from $\mathcal{U}[0,A_{\max}]$.

The model is described by seven parameters:
\begin{itemize}
    \item $\mu_0$ rules the number $\mu_s$ of spontaneous initial pulses per GRB, which is sampled from a Poisson distribution with $\mu_0$ as expected value:
    \begin{equation} 
        p(\mu_s|\mu_0) = \frac{\mu_0^{\mu_s} \exp(-\mu_0)}{\mu_s!}\quad.
    \end{equation} 
    \item $\mu$ rules the number of child pulses $\mu_c$ generated by each parent pulse, which is sampled from a Poisson distribution with $\mu$ as expected value:
    \begin{equation} 
        p (\mu_c|\mu) = \frac{\mu^{\mu_c} \exp(-\mu)}{\mu_c!} \quad.
    \end{equation}
    \item $\alpha$ rules the delay $\Delta t$ between a child and its parent. This delay is exponentially distributed, with e-folding time given by $(\alpha\tau)$, where $\tau$ is the time constant of the child pulse:
    \begin{equation} 
        p(\Delta t) = (\alpha \tau)^{-1} \exp(-\Delta t /\alpha \tau)\quad.
    \end{equation}
    Moreover, the spontaneous $\mu_s$ primary pulses are all assumed to be delayed with respect to a common invisible trigger event; the probability distribution of such delay $t$ is exponentially distributed:
    \begin{equation} 
        p(t) = (\alpha \tau_0)^{-1} \exp(- t /\alpha \tau_0)\quad,
    \end{equation}
    $\tau_0$ being the time constant of the primary pulse.
    \item $\tau_{\min}$ and $\tau_{\max}$ define the boundaries for the constant $\tau_0$ of the initial spontaneous pulses and whose probability density function is $p(\tau_0) \propto 1/\tau_0$, equivalent to a uniform distribution of $\log \tau_0$:
    \begin{equation} 
        p (\log \tau_0) = \left[\log \tau_{\max}-\log \tau_{\min}\right]^{-1}\quad,
    \end{equation}
    where $\tau_{\min}$ has to be shorter than the time resolution of the instrument. Varying $\tau_{\max}$ is equivalent to rescaling all average avalanche properties in time.
    \item $\delta_1$ and $\delta_2$ define the boundaries, $[\delta_1, \delta_2]$, of a uniform distribution assumed for the logarithm of the ratio between $\tau$ of the child and $\tau_p$ of its parent:
    \begin{equation} 
        p[\log (\tau/\tau_p)] = \left|\delta_2 - \delta_1\right|^{-1}\quad,
    \end{equation}
    with $\delta_1 < 0$, $\delta_2 \geq 0$, and $|\delta_1| > |\delta_2|$.
\end{itemize}
Each of the $\mu_s$ spontaneous initial pulses gives rise to a pulse avalanche, acting as a parent, spawning another set of child pulses, in a recurrent way. The avalanche ends when it converges due to subcritical values of the model parameters. Finally, it is the superposition of the parents and all the generations of child pulses created during the avalanche that shapes the LC of an individual GRB.

The stochastic pulse avalanche model was used to simulate both BATSE and {\it Swift}/BAT LCs. The statistical noise depends on the total counts in each time bin, which requires the knowledge of the typical background count rate for a given instrument. For BATSE, which consisted of NaI(Tl) scintillators, we assumed a constant background rate of 2.9 cnt $\si{s^{-1} cm^{-2}}$, which corresponds to the median of the distribution of the measured error rates. Each final simulated LC was the result of a Poisson realisation, assuming for each time bin the total counts (that is, noise-free simulated profile plus background) as expected value. Lastly, the background was removed.

{\it Swift}/BAT is a coded mask coupled with a CZT detection array. Its background-subtracted LCs are the result of the deconvolution of the detection with the pattern of the mask, so the rate in each time bin can be modelled as a Gaussian variable. To simulate BAT LCs, the rate of each time bin was sampled from a Gaussian distribution centred on the LC (noise-free) model obtained with the pulse avalanche model, and with standard deviation randomly sampled from the errors measured in the real {\it Swift}/BAT LCs.

All the simulations were carried out using an open-source Python\footnote{\url{https://www.python.org/}} implementation\footnote{\url{https://github.com/anastasia-tsvetkova/lc_pulse_avalanche}} by one of the authors.

\subsection{Genetic algorithm}
\citetalias{SternSvensson96} proposed a set of values for the seven model parameters as the result of an educated guess. The optimisation of these parameters, however, is an ideal task for nowadays routinely used ML techniques.

GAs are a specific type of algorithms in the larger family of the so-called evolutionary algorithms~\citep{russell2021artificial, rojas1996, aggarwal2021, hurbans2020}, where a Darwinian evolution process is simulated to find the parameters that maximise a function.

In GAs, each solution to an optimisation problem can be seen as an \textit{individual}, with the ``fitness'' of that individual being determined by the objective function value of the corresponding solution. These solutions are points in the domain of the function to be optimised. In our work, each individual is represented by a genome made of seven genes, which are the parameters of the~\citetalias{SternSvensson96} model described in Section~\ref{sec:ss96model}.

At each generation, a new set of individuals is created. Over time, the points belonging to the new generations gradually converge towards local maxima of the fitness function. In order to improve over successive generations the overall fitness of the population, GAs incorporate three fundamental processes: selection, crossover, and mutation.

The typical life cycle of a GA, made up of a succession of the so-called \textit{generations}, includes the following steps:
\begin{enumerate}
    \item Population initialisation: Generating randomly a population of potential solutions;
    \item Evaluating fitness: Assessing the quality of each individual by employing a fitness function that assigns scores to evaluate their fitness;
    \item Parent selection: Choosing pairs of parents for reproduction based on their fitness score;
    \item Offspring creation: Producing offspring by combining genetic information from parents, and introducing random mutations;
    \item Generation advancement: Selecting individuals and offspring from the population to progress to the next generation.
\end{enumerate}

GAs are particularly useful in situations where there is no available information about the function's gradient at the evaluated points. Indeed, GA can effectively handle functions that are not continuous or differentiable~\citep{rojas1996}.

\subsection{Parameter optimisation}
The GA has been implemented using \texttt{PyGAD}\footnote{\url{https://github.com/ahmedfgad/GeneticAlgorithmPython}}, an open-source Python library containing a collection of several ML algorithms~\citep{gad2023pygad}.

We constrain the seven parameters of the model within the intervals shown in Table~\ref{table::parameterrange}. 

\begin{table}
\caption{Region of exploration during the GA optimisation of the seven parameters of the~\citetalias{SternSvensson96} stochastic model.} 
\label{table::parameterrange} 
\centering 
    \begin{tabular}{c c c}
    \hline\hline
    Parameter      & Lower bound    & Upper bound \\ 
    \hline
    $\mu$          & $0.80$         & $1.7$ \\ 
    $\mu_0$        & $0.80$         & $1.7$ \\
    $\alpha$       & $1$            & $15$ \\
    $\delta_1$     & $-1.5$         & $-0.30$ \\
    $\delta_2$     & $0$            & $0.30$ \\
    $\tau_{\min}$  & $0.01\,\si{s}$ & $\texttt{bin\_time}\,\si{s}$ \\
    $\tau_{\max}$  & $1\,\si{s}$    & $60\,\si{s}$ \\ 
    \hline 
    \end{tabular}
\end{table}

The GA evolves through a sequence of generations consisting of a population with $N_{\rm pop} = 2000$ individual sets of seven parameters. Each set of parameters, hereafter referred to as an individual, is then used to generate $N_{\rm grb} = 2000$ LCs. The very same three constraints, mentioned in Section~\ref{ss:lc_selection}, and used for the selection of BATSE and {\it Swift}/BAT dataset, are applied also on the simulated GRB LCs, the generated ones not satisfying such constraints being discarded. 
For each of the $N_{\rm pop}$ individuals, we evaluate the same four metrics defined in~\citetalias{SternSvensson96} over the corresponding $N_{\rm grb}$ LCs (cfr.~Section~\ref{sec:metrics}), and compare them with the values obtained from the real datasets, by computing the $L2$ loss between these four observables. The final loss associated with a given individual is simply defined as the average of these four quantities, the fitness score being the inverse of this value.

Individuals are then ranked based on their loss. The next generation of individuals is obtained by mixing the genes (i.e.~the values of the seven parameters) of the fittest individuals in the current generation. No individuals are instead automatically kept in the next generation, that is, we set to zero the so-called ``elitism''\footnote{Due to the stochastic nature of the~\citetalias{SternSvensson96} algorithm, the same set of parameters will never produce a set of LCs with the same loss; therefore keeping a set of individuals in the next generation is not helping, since in reality, given seven fixed parameters, there are fluctuations in the value of the corresponding loss.}.
The offspring is obtained by randomly sampling two individuals among the top $15\%$ in the current generation and assigning to each gene the value of the seven parameters from one of the two parents, with equal probability. 

Finally, we include the possibility for genetic random mutations to occur. During the mating step, each one of the seven parameters has a $4\%$ probability of undergoing mutation, meaning that the value of the parameters is not inherited from one of the two parents, but instead, it is randomly sampled from the exploration range of the parameter (Table~\ref{table::parameterrange}).

The optimisation process is stopped when convergence of the loss, and thus of the value of the seven parameters, is reached.


\section{Results}
\label{sec:res}
In Table~\ref{table::parameterresults} we compare the values of the seven model parameters suggested in~\citetalias{SternSvensson96} with the results of our GA optimisation on the BATSE and \textit{Swift} training datasets. The final optimised values of the seven parameters are obtained as the median value in the whole population of the last GA generation. We also list the achieved values of the loss function evaluated on the training set (both in terms of best parameter configuration and by averaging on the last population) as well as on the test set (i.e. estimated by using 5000 newly simulated GRB LCs). In the bottom, we resolve the individual contribution of each component to the test loss.
\begingroup
\setlength{\tabcolsep}{6pt}       
\renewcommand{\arraystretch}{1.4} 
\begin{table}
\caption{Results of the GA optimisation on the BATSE and \textit{Swift} datasets.} 
\label{table::parameterresults}
\centering 
    \begin{tabular}{c c c c}
    \hline\hline
    Parameter & \citetalias{SternSvensson96} & BATSE & \textit{Swift}/BAT\\ 
    \hline
    $\mu$            & $1.20$         & $1.10^{+0.03}_{-0.02}$         & $1.34^{+0.03}_{-0.02}$\\ 
    $\mu_0$          & $1.00$         & $0.91^{+0.06}_{-0.07}$         & $1.16^{+0.18}_{-0.10}$\\ 
    $\alpha$         & $4.00$         & $2.57^{+0.07}_{-0.52}$         & $2.53^{+0.25}_{-0.01}$\\ 
    $\delta_1$       & $-0.50$        & $-1.28^{+0.16}_{-0.05}$        & $-0.75^{+0.11}_{-0.29}$\\ 
    $\delta_2$       & $0$            & $0.28^{+0.01}_{-0.03}$         & $0.27^{+0.01}_{-0.02}$\\ 
    $\tau_{\min}$    & $0.02\,\si{s}$ & $0.02^{+0.02}_{-0.01}\,\si{s}$ & $0.03^{+0.02}_{-0.02}\,\si{s}$\\ 
    $\tau_{\max}$    & $26.0\,\si{s}$ & $40.2^{+0.9}_{-1.2}\,\si{s}$   & $56.8^{+0.4}_{-1.3}\,\si{s}$\\
    \hline 
    Loss (\textit{Train} best)                          & --      & $0.72$ & $0.38$\\
    Loss (\textit{Train} avg.)                          & --      & $0.98$ & $0.66$\\
    \hline 
    Loss (\textit{Test})                                & $1.47$  & $0.88$ & $0.56$\\
    \hline 
    Loss (\textit{Test}: $\langle F / F_p\rangle$)      & $1.01$  & $0.67$ & $0.46$\\
    Loss (\textit{Test}: $\langle(F / F_p)^3\rangle$)   & $0.40$  & $0.20$ & $0.20$\\
    Loss (\textit{Test}: $\langle\mathrm{ACF}\rangle$)  & $2.24$  & $0.64$ & $0.49$\\
    Loss (\textit{Test}: $T_{20\%}$)                    & $2.22$  & $2.04$ & $1.08$\\
    \hline 
    \end{tabular}
    \tablefoot{Col.~2 presents the parameters given by~\citetalias{SternSvensson96} (for the BATSE dataset), while Col.~3 and Col.~4 show the optimised ones obtained after 30 generations of the GA for BATSE and {\it Swift}/BAT, respectively. From the distribution of the seven parameters in the last generation we estimated their best-fitting values as the median, and their corresponding errors as the 16-th and 84-th percentiles. ``Train best'' is the loss of the best generation, while ``Train avg.'' is the average loss in the last generation. The test set is a newly produced set of 5000 simulated LCs; the last four rows show all the single contributions to the ``Test'' loss.
    }
\end{table}
\endgroup

\begin{figure*}
   \centering
   \includegraphics[width=0.85\linewidth]{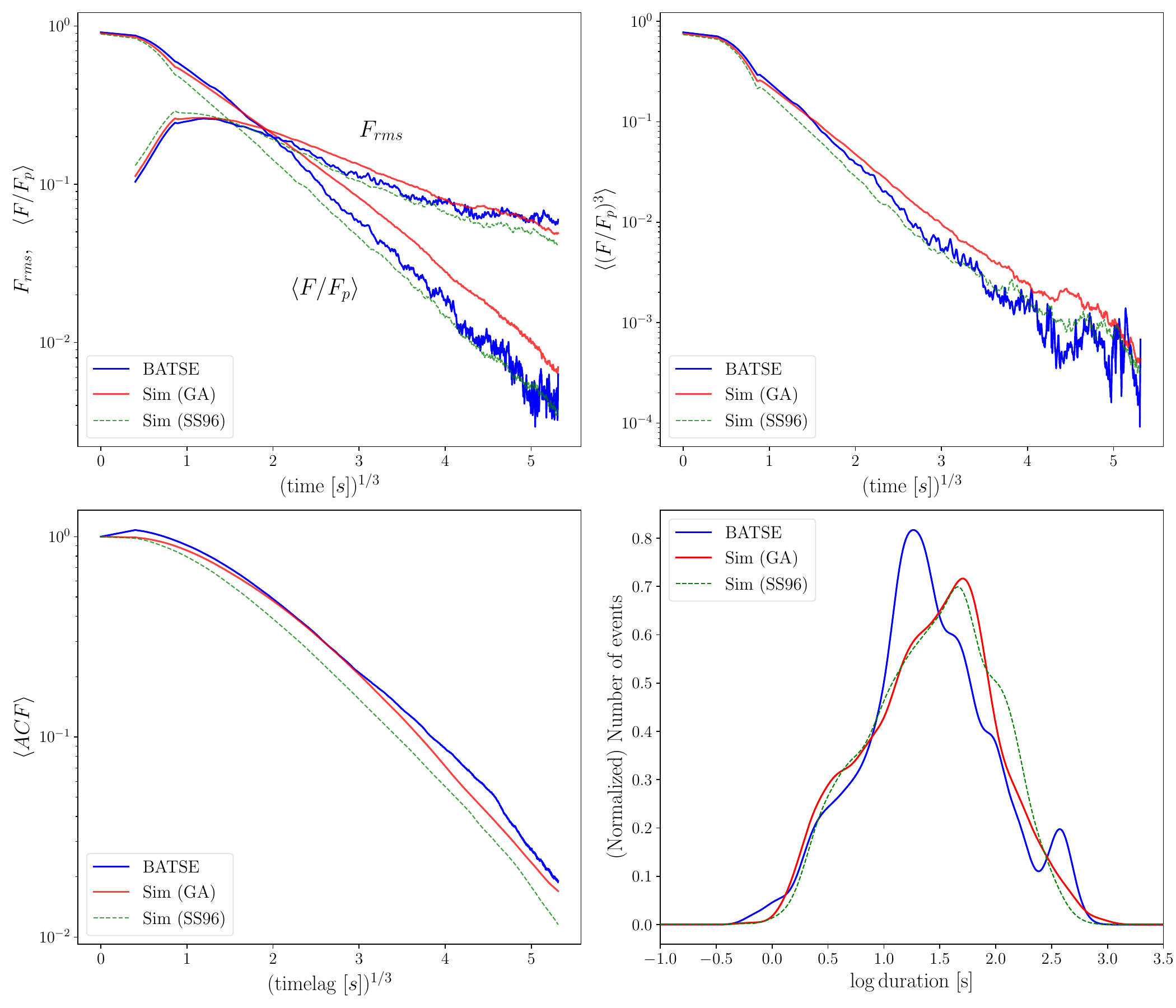}
   \caption{Average distributions of real (blue), simulated GA-optimised (red), and simulated~\citetalias{SternSvensson96} (green) BATSE GRB profiles, estimated on the test set (see Table~\ref{table::parameterresults}). \textit{Top left}: average peak-aligned post-peak normalised time profile, together with the r.m.s. deviation of the individual peak-aligned time profiles, $F_{\mathrm{rms}} \equiv\big[\langle(F / F_p)^2\rangle-\langle F / F_p\rangle^2\big]^{1/2}$. \textit{Top right}: average peak-aligned third moment test. \textit{Bottom left}: Average ACF of the GRBs. \textit{Bottom right}: distribution of duration, measured at a level of $20\%$ of the peak amplitude ($T_{20\%}$). In \textit{top left} and \textit{top right} panels, both real and simulated averaged curves were smoothed with a Savitzky-Golay filter to reduce the effect of Poisson noise. In \textit{bottom right} panel, a Gaussian kernel convolution has been applied to both real and simulated distributions.
   }
   \label{fig::4observables_L_tot_4_batse_comparison}
\end{figure*}

\begin{figure*}
   \centering
   \includegraphics[width=0.85\linewidth]{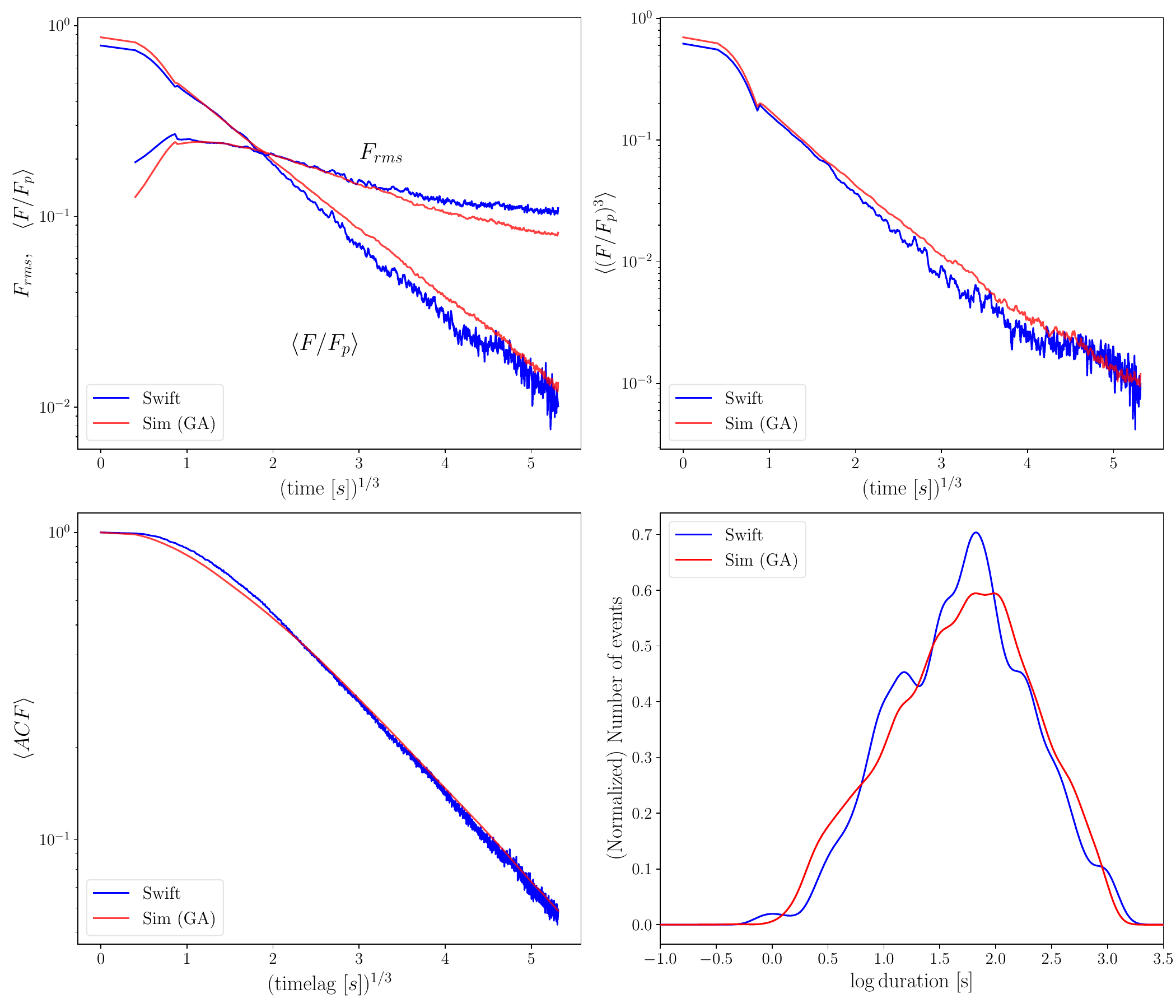}
   \caption{Comparison between the real \textit{Swift}/BAT dataset and the corresponding simulated dataset on the same four metrics defined for the BATSE dataset, analogously to Figure~\ref{fig::4observables_L_tot_4_batse_comparison}.
   }
   \label{fig::4observables_L_tot_4_swift}
\end{figure*}

Figure~\ref{fig::4observables_L_tot_4_batse_comparison} displays the comparison of the four observables, described in Section~\ref{sec:metrics}, between the real BATSE curves and simulated ones (test set). In particular, the panels show the average profiles obtained from the 585 useful BATSE events (blue), the ones estimated from 5000 simulated GRBs with optimised parameters (red), and the ones estimated from 5000 LCs simulated using the parameter values guessed by~\citetalias{SternSvensson96} (green).
Figure~\ref{fig::4observables_L_tot_4_swift} shows the analogous comparison between the simulated and the real {\it Swift}/BAT curves.

We find an excellent agreement for three out of the four metrics computed from real and simulated BATSE LCs, in particular for the average post-peak time profile, its third moment, and the average auto-correlation, whose $L2$ loss values are smaller than the corresponding ones estimated with~\citetalias{SternSvensson96} non-optimised parameters, as can be seen from the bottom part of Table~\ref{table::parameterresults}. For instance, the average ACF metric shows a relative improvement of $\sim74\%$ after the optimisation. The $T_{20\%}$ distribution holds the largest contribution to the loss; yet, it slightly improves the~\citetalias{SternSvensson96} performance ($\sim8\%$ relative improvement). Overall, compared with~\citetalias{SternSvensson96}, our GA-optimised results on BATSE data better reproduce the observed distributions. 

As can be inferred from Table~\ref{table::parameterresults}, and graphically from Figure~\ref{fig::4observables_L_tot_4_swift}, according to the loss function the avalanche model appears to work even better in the case of {\it Swift} data: the results on the average ACF and third moment of peak-aligned profiles are comparably good, whereas the average peak-aligned profile and duration distributions are significantly improved with respect to the BATSE case, with a relative loss decrease of $\sim43\%$ and $\sim58\%$, respectively.

Notably, the two sets of best-fitting parameters obtained with BATSE and with {\it Swift}/BAT are very similar and, surprisingly, overall not too different from the one guessed by~\citetalias{SternSvensson96}. In Section~\ref{sec:disc_conc} we discuss the relevance of this result in more detail.

The parameters for which our optimal values for both sets are somewhat different from those of~\citetalias{SternSvensson96} are $(\delta_1, \delta_2)$, $\tau_{\max}$, and $\alpha$. The former pair defines the dynamic range of the child-to-parent pulse duration ratio: our values turn into broader dynamical ranges than~\citetalias{SternSvensson96}, and, at variance with those authors, they admit the possibility of children lasting longer than parents, being $\delta_2>0$. 
While~\citetalias{SternSvensson96} assumed $\tau_{\max}=26$~s as the maximum value for the duration of parent pulses, our optimised solution favours the possibility of longer parent pulses: $40.2$~s for BATSE and $56.8$~s for {\it Swift}/BAT.
Finally, our best-fit values for $\alpha$, which rules the time delay between parent and child, lean towards slightly shorter intervals than~\citetalias{SternSvensson96}.

\begin{figure*}
    \centering
    \includegraphics[width = 0.95\textwidth]{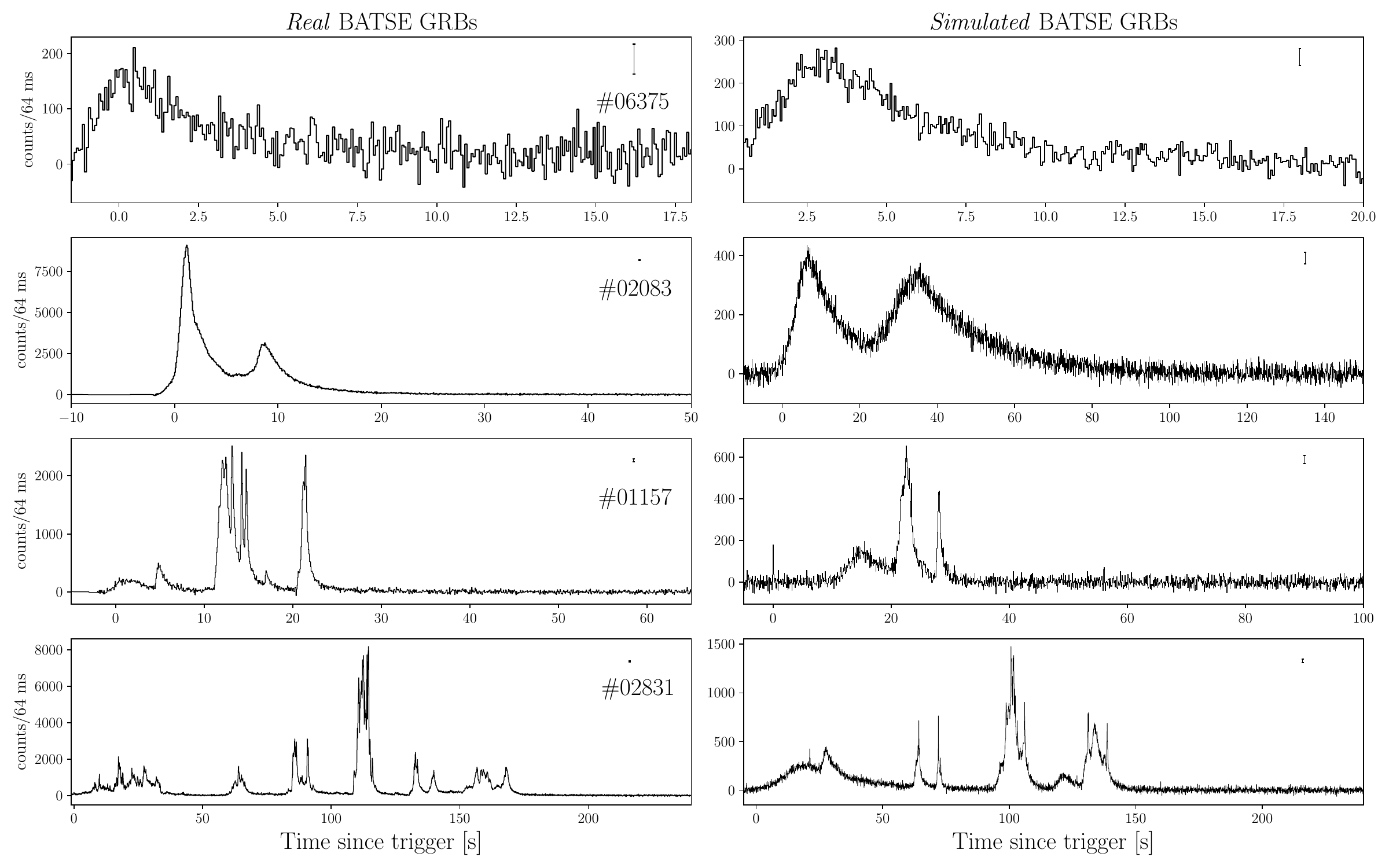}
    \caption{Four examples of as many classes of GRB LCs from the BATSE real sample ({\it left}) along with their trigger number, and the corresponding simulated one ({\it right}). Following the same qualitative classification adopted by~\citetalias{SternSvensson96}, from top to bottom the four classes are ``single pulse'', ``blending of some pulses'', ``moderately structured'', and ``highly erratic''. On the top right of each subplot is shown the average error on the counts of the corresponding LC.} 
    \label{fig::fig-real-fake-paper}
\end{figure*}

As in~\citetalias{SternSvensson96}, Figure~\ref{fig::fig-real-fake-paper} presents the comparison of four real BATSE time profiles with four simulated LCs of similar morphology and complexity, sampled from the test set, generated using the best-fitting set of BATSE model parameters given above. 
This qualitative plot shows the ability of the~\citetalias{SternSvensson96} stochastic model to reproduce the different morphological classes of LCs discussed in the literature from the earliest observations~\citep{Fishman1995}.


\section{Discussion and conclusions}
\label{sec:disc_conc}

For the first time in the GRB literature, here we developed and implemented an ML technique to optimise the parameters of a stochastic model capable of generating ex-novo realistic GRB LCs. Our work confirmed the soundness of the insight by~\citetalias{SternSvensson96}: a simple toy model like the stochastic pulse avalanche one is able to generate populations of LCs, whose average behaviour closely resembles that of the real populations of BATSE and {\it Swift}/BAT long GRBs. With the implementation of GA, we found the two best fit sets of the seven parameters of the model that best reproduce the average behaviours of the two datasets, thus (i) progressing from the educated-guessed values of the original paper to a real fit of the model on BATSE data, and (ii) applying it for the first time to an independent catalogue of GRB LCs like {\it Swift}/BAT, whose data differ from the BATSE one in many aspects, as detailed below.

In light of our GA-optimised results on BATSE data, the educated guess by~\citetalias{SternSvensson96} turns out to be surprisingly good. In particular, the finding that $\mu$, that is the average number of child-pulses generated by each parent-pulse, must be close to unity (our $1.10_{-0.02}^{+0.03}$ vs. $1.20$ of~\citetalias{SternSvensson96}) confirms the insightful Ansatz by~\citetalias{SternSvensson96} that the model must operate very close to a critical regime ($\mu=1$), to account for the observed variety of GRB profiles. Interestingly, the same clue is also obtained in the GA-optimised parameters of the {\it Swift}/BAT sample.
This result is far from obvious for three main reasons: (i) the passband of the two experiments is significantly different, with {\it Swift}/BAT profiles being softer and, as such, less spiky (e.g., \citealt{Fenimore95}); (ii) the average S/N of the two sets is also remarkably different, with a minimum value of 70 for BATSE, to be compared with the poorer lower threshold of 15 for {\it Swift}/BAT (see Section~\ref{sec:data}); (iii) the GRB populations seen by the two experiments are likely different: thanks to its larger effective area at low energies, longer trigger accumulation times and much more complex trigger algorithms, {\it Swift}/BAT detects more high-redshift GRBs~\citep{Band06, Lien14, WandermanPiran10}. Therefore, our results provide additional evidence for a nearly critical regime in which GRB engines would work, in agreement with other independent investigations~\citep{Maccary24, Guidorzi24}.

Within models of GRB prompt emission that ascribe the observed variability to the GRB inner engine, such as internal shocks (IS; \citealt{Rees94, Kobayashi97, Daigne98, Maxham09}), the avalanche character of the stochastic model would directly reveal the way the newborn compact object releases its energy over time. If this is an hyperaccreting BH, magneto-rotational and/or gravitational instabilities in the accretion disc could cause fragmentation, as proposed to explain GRB X-ray flares~\citep{Perna06}, which are also known to have an internal origin and share other properties with the prompt emission pulses~\citep{Burrows05b, Chincarini10, Maxham09, Guidorzi15b}. This could explain the branching character of the avalanche process, where child pulses are the result of further fragmentation from parent episodes. Analogously, viscous instabilities could lead to the formation of clumps~\citep{Kawanata13, Shahamat21}, whose accretion might develop like a branching process.

An important role in driving the kinds of instability that finally trigger GRB prompt emission, is likely played by magnetic fields.
In the ICMART model \citep{ICMART}, which builds upon the idea of IS between magnetised shells, prompt emission is powered by dissipation of magnetic energy through a runaway sequence of reconnection events triggered by a so-called ICMART event. The latter would result from the progressive distortion of initially ordered magnetic fields caused by a sequence of IS of magnetised shells. The runaway character of the energy release could in principle end up in a branching mechanism of elementary episodes, such as those envisaged in the~\citetalias{SternSvensson96} model.

In addition to providing new clues on the dynamical behaviour of LGRB inner engines or, more generally, on the way some kind of energy is dissipated into gamma-rays, this model offers the practical possibility of simulating realistic GRB profiles with future experiments, such as HERMES~\citep{Fiore20_HERMES} and possibly the X/Gamma-ray Imaging Spectrometer (XGIS; \citealt{Amati22}) aboard ESA/M7 candidate THESEUS~\citep{Amati21b} currently selected for a phase A study. The task of simulating realistic GRB time profiles, as they will be seen by forthcoming detectors, is far from obvious: the alternative option of renormalising real LCs observed with different instruments is inevitably hampered by the presence of counting statistics (Poisson) noise, which cannot be merely rescaled without altering its nature. A filtering procedure would be then required, which in turn assumes that the uncorrelated Poisson noise can be disentangled from the genuine (unknown) variance of GRBs, which also requires substantial effort.

Summing up, the present work showcases the potential of a simple toy model like the avalanche one conceived by~\citetalias{SternSvensson96}, once it is properly bolstered with ML techniques. Moreover, it paves the way to further optimisation of the model in different directions: (i) by adding further metrics, such as the distributions of the following observables: GRB S/N, duration of observed pulses, or the number of peaks per GRB~\citep{Guidorzi24}; (ii) by studying in more detail the dependence of the model parameters on the energy channels; (iii) by carrying out the same study in the comoving frame of a sample of GRBs with known redshift, assuming the luminosity and released energy distributions of individual pulses~\citep{Maccary24}. Eventually, these efforts should end up with a reliable and accessible machine for simulating credible LGRB profiles with any experiment. In parallel, the final outcome would be a detailed characterisation of the dynamics that rules long GRB prompt emission, possibly disclosing the nature of long GRB engines.

The source code of our algorithm, alongside all the scripts used to perform the data analysis and produce the plots, is publicly available on GitHub\footnote{\url{https://github.com/LBasz/geneticgrbs/tree/arxiv-v1}}.


\begin{acknowledgements}
LB is indebted to the communities behind the multiple free, libre, and open-source software packages on which we all depend. AT acknowledges financial support from ASI-INAF Accordo Attuativo HERMES Pathfinder operazioni n. 2022-25-HH.0 and the basic funding program of the Ioffe Institute FFUG-2024-0002. \\
CG conceived the research based on the original code developed by AT according to the SS96; LB, GA, and CG developed the methodology; LB, LF and CG performed all the numerical work, including software development, investigation, and validation. All authors contributed to the discussion of the results, the editing and revision of the paper.\\
We thank the anonymous referee for the helpful comments that improved the paper.\\
This work uses the following software packages:
    \href{https://www.python.org/}{\texttt{Python}}
    \citep{python},
    \href{https://github.com/ahmedfgad/GeneticAlgorithmPython}{\texttt{PyGAD}}
    \citep{gad2023pygad},
    \href{https://github.com/numpy/numpy}{\texttt{NumPy}}
    \citep{numpy1, numpy2},
    \href{https://github.com/scipy/scipy}{\texttt{SciPy}}
    \citep{scipy},
    \href{https://github.com/matplotlib/matplotlib}{\texttt{matplotlib}}
    \citep{matplotlib},
    \href{https://github.com/mwaskom/seaborn}{\texttt{seaborn}}
    \citep{seaborn},
    \href{https://www.fe.infn.it/u/guidorzi/new_guidorzi_files/code.html}{\sc{mepsa}}
    \citep{guidorzi2015mepsa},
    \href{http://www.gnuplot.info/}{\texttt{gnuplot}}
    \citep{gnuplot6.0},
    \href{https://www.gnu.org/software/bash/}{\texttt{bash}}
    \citep{gnu2007free}.
    
\end{acknowledgements}


%
%
\bibliographystyle{aa}
\bibliography{bibliography,software,alles_grbs}
\end{document}